\documentstyle[newpasp,twoside,epsf]{article}
\newcommand{\rashort}[2]{$#1^{\mathrm{h}}#2^{\mathrm{m}}$}

\newcommand{\etal}{{et al.}\/ }

\begin{document}

\title{The cluster environments of radio loud quasars}

\author{J. M. Barr, M. N. Bremer}
\affil{Department of Physics, University of Bristol, Tyndall Ave, Bristol BS8 1TL, U.K.}

\author{J. C. Baker}
\affil{Astronomy Department, University of California, Berkeley, U.S.A.}

\author{M. D. Lehnert}
\affil{Max-Planck-Institut f\"ur extraterrestrische Physik, Garching, Germany}

\begin{abstract}

We have carried out multi-colour imaging of the fields of a
statistically complete sample of low-frequency selected radio loud
quasars at $0.6<z<1.1$, in order to determine the characteristics of
their environments.  The largest radio sources are located in the
field, and smaller steep-spectrum sources are more likely to be found
in richer environments, from compact groups through to clusters. This
radio-based selection (including source size) of high redshift groups
and clusters is a highly efficient method of detecting rich
environments at these redshifts. Although our single filter clustering
measures agree with those of other workers, we show that these
statistics cannot be used reliably on fields individually, colour
information is required for this.
 
\end{abstract}

\section{Introduction}

Powerful radio sources at low redshift, (Cyg A, 3C295), are known to
reside in massive elliptical galaxies such as those commonly found at
the centre of clusters. AGN at low redshift have been shown to
exist in environments of above average galactic density (Yee \& Green
1987) and studies extending these measurements to higher redshift show
no evolution in this trend (Ellingson, Yee, \& Green 1991; Hill \&
Lilly 1991; Yee \& Ellingson 1993; Wold et al. 2000). It is therefore
reasonable to expect that at high redshift we may be able to find
clusters by looking at the environments of AGN. This is further
substantiated by theoretical arguments regarding the formation of {\it
extended} lobe-dominated radio structures ({\it e.g.} Miley
1980). Such models require a confining medium to exist over hundreds
of kpc, the scale lengths and pressures being comparable with those
found in clusters of galaxies. An advantage of instigating a search in
this manner is provided by the fact that a quasar will pinpoint the
redshift of interest and through careful choice of filters facilitate
the identification of associated galaxies by way of their colours.

These reasons make it possible to build a sample of distant groups and
clusters from a study of radio loud quasars (RLQs), although care must
be taken to understand the selection function. With reasonable
estimates regarding the lifetime of the AGN and its effect on the
cluster this can be achieved. In any case the opportunity to study
clusters and their constituents is not lost. A related benefit is that
we may find and study small and compact groups at high
redshift. Current optical and X-ray methods will be insensitive to
such low mass/luminosity systems and the environments of RLQs may
offer the best method of finding significant numbers of these systems
at high redshift for some time to come.

\section{Sample and observations}

The quasars used for this study are drawn from the Molonglo quasar
sample (MQS, Kapahi et al. 1998) of low frequency selected radio loud
quasars. This is a complete sample of 111 RLQs with S$_{408} > 0.95$
Jy in a contiguous area of southern sky ($-30^{\circ} > \delta >
-20^{\circ}$, but excluding the R.A.  range \rashort{14}{03} -
\rashort{20}{03}).  


Within the range \rashort{0}{0}$< \mathrm{RA} <$\rashort{14}{0} there
are 30 MQS objects which have $0.6<z<1.1$, 22 for which we have
data. Eight sources with these criteria were {\em randomly} excluded
due to observing time restrictions.  This sample has many
merits. Firstly it has a single selection criterion, that of low
frequency radio luminosity. Thus it does not attempt to ``match''
catalogues of what may be intrinsically different objects. Secondly,
of all objects in our chosen redshift range none are excluded based on
optical or radio bias. The remaining objects can therefore be thought
of as a {\em representative} sample. Also, the low frequency selection
avoids preferentially selecting core dominated objects, ie: those
which may be of intrinsically lower luminosity but have Doppler
boosted emission due to their orientation to the line of sight.  The
MRQ is thus useful as a ``complete and unbiased'' sample on which to
study the environmental properties of RLQs.

We have observed our objects in 2 or more filters which straddle the
redshifted 4000\AA \ break in order to maximise the contrast for early
type galaxies. In several cases we also have near infrared (NIR) $J$
and $K'$ filters. The telescopes used were the CTIO 4m, the ESO 2.2m
and 3.6m and the AAT.

\section{Single filter analysis}

\begin{figure}

 	\plotfiddle{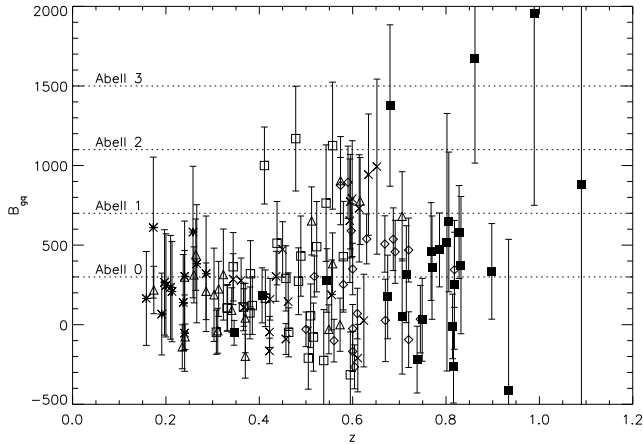}{4.5cm}{0}{50}{50}{-144}{-25}

	\vspace{0.5cm}

	\caption{ \small{Values of the spatial cross correlation amplitude as
  a function of redshift for radio loud quasar fields. Filled squares,
  this work. Diamonds, Wold \etal (2000). Triangles, Yee \& Ellingson
  (1993). Squares, Ellingson \etal (1991). Crosses, Yee \& Green
  (1987). Asterisks, McClure \& Dunlop (2000). The horizontal lines
  represent the Abell classes quoted in McClure \& Dunlop.}}

\end{figure}

Previous investigators have sought to quantify the excess of objects
around quasars, (or radio galaxies, or any other object), by counting
sources within a certain radius in a single filter (0.5 Mpc in most
cases). We initially analyse clustering around quasars with reference
to such a measure, the galaxy-quasar spatial cross correlation
amplitude, $B_{gq}$. A detailed derivation can be found in Yee \&
Green (1984) and Longair \& Seldner (1979).

The results of applying this method to our data can be seen in figure
1, as well as those for other investigators. Our results confirm that
the quasars occupy a variety of environments and show no evidence for
evolution to a redshift of 1.

There are, however, reasons why these methods are not necessarily the
most accurate to use at high redshift. Firstly, they assume that the
AGN is at the centre of any (roughly circular) overdensity. As we
progress to higher redshifts, clusters are less likely to be in
relaxed spheroidal systems and may exhibit significant substructure
which can extend further than 0.5 Mpc from the quasar.  Secondly, the
number density of objects at the magnitude levels of interest displays
variation on the scale-lengths of distant clusters making it difficult
to estimate the background to subtract from any overdensity.  Thirdly,
even when an obvious overdensity is discovered, it may consist of
objects which have the wrong colours to be galaxies at the redshift of
the quasar (see fig. 2b). In this case these statistical analyses will
not only be wrong, they may bias the average result systematically and
in doing so render any measure of the average properties of a sample
incorrect.

\section{Multi-colour results}

\begin{figure}

	\plotfiddle{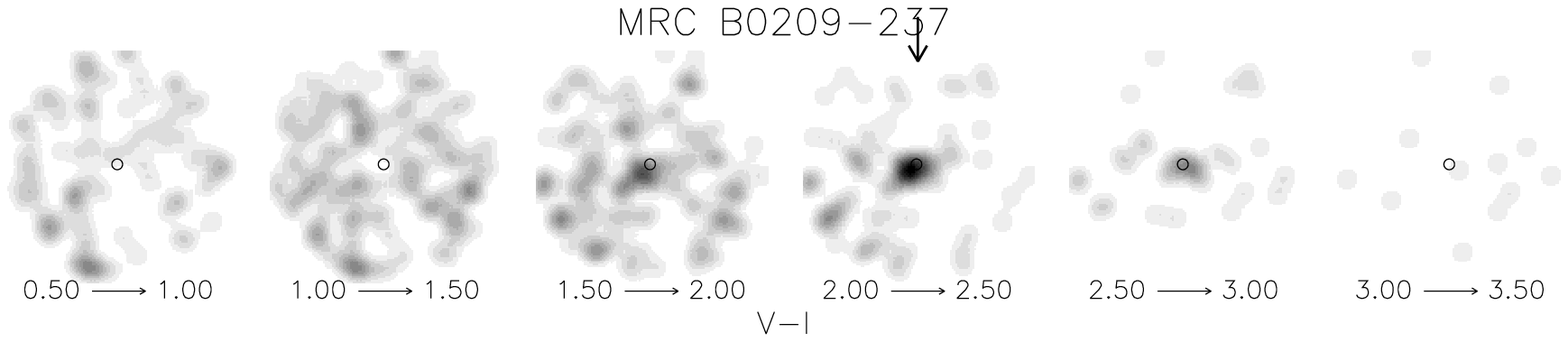}{1cm}{0}{60}{60}{-150}{-20}

	\plotfiddle{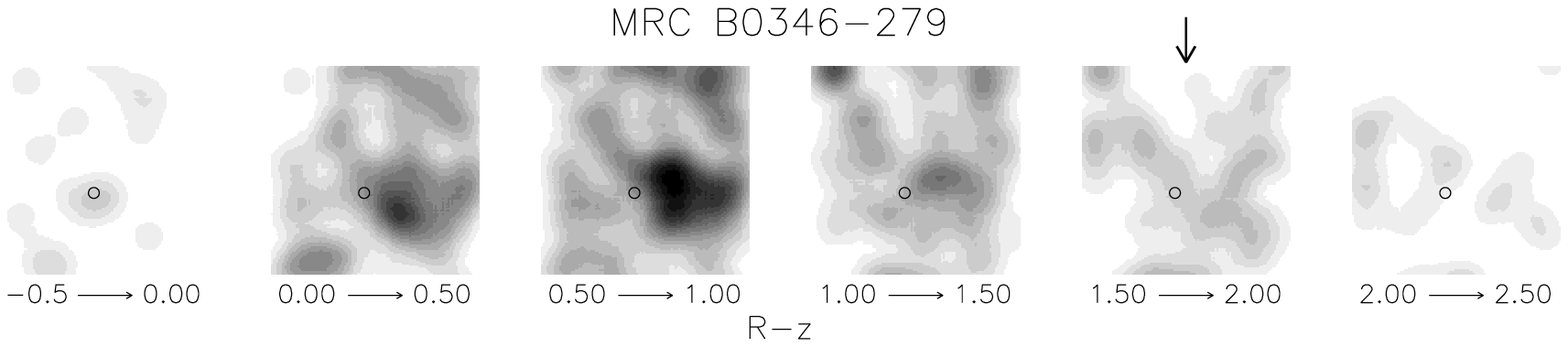}{1cm}{0}{60}{60}{-150}{-50}

	\plotfiddle{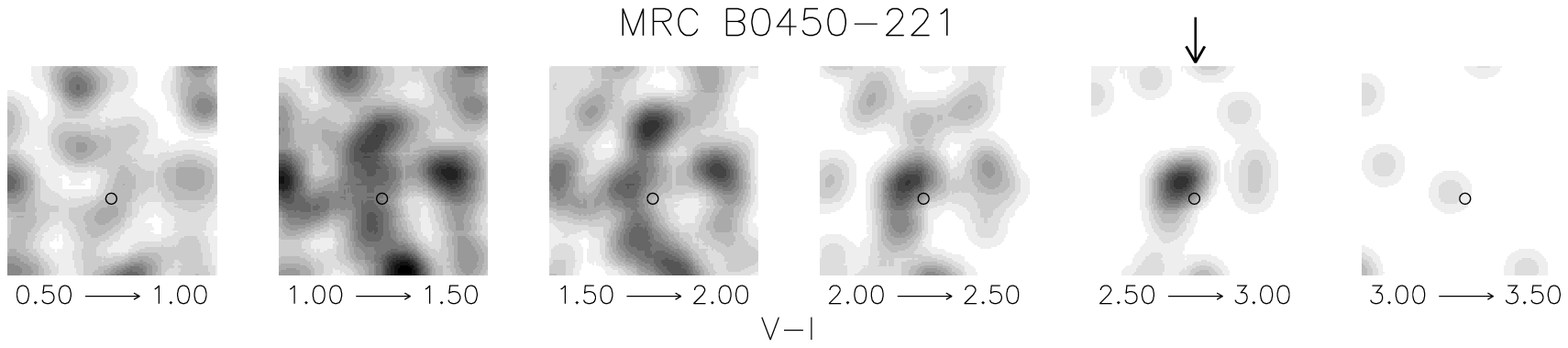}{1cm}{0}{60}{60}{-150}{-80}
        
	\vspace{2.5cm}

  	\caption{ {\small The surface density of objects as a function
  of position in the fields of three of our quasars, binned according
  to colour range. Cluster cores at the redshift of the quasar would
  produce peaks in the density maps indicated by the arrow. While all
  three fields produce high values of $B_{gq}$, we can see that in the
  second case the overdensity is at the wrong colour for the quasar
  redshift. The position of the quasar is marked by the circle.}}

\end{figure}

By choosing optical filters which straddle the 4000\AA \ break, we can
select the passively evolving elliptical galaxies at the redshift of
the quasar. Figure 2. shows the density of objects around three of our
quasars separated according to colour. We can identify agglomerations
as having the potential to be at the quasars redshift by using these
sort of plots. Further analysis looking at the colour dispersion or
for a red sequence or using NIR filters can then quantify accurately
which galaxies belong to the cluster core and which do not. Examples
of these techniques are to be found in Baker et al. (2001) and Bremer
et al. (2001).

Because of the arguments regarding the formation of the radio lobes
being dependent on a large scale medium, we would expect to see a
correlation between the size of the source and the clustering around
it. We see just such a correlation in figure 3. Our sample is
separated into those sources with a radio size $<20''$ and those with
$>20''$. Unresolved flat spectrum sources, ($\alpha < 0.6$), are
rejected in order to prevent beamed sources with intrinsically lower
luminosity from being included. The remaining sample of 19 steep
spectrum objects is thus split approximately in half. If we loosely
define our clusters as those with more than 10 associated core
members, we see that five out of our six clusters are associated with
smaller radio structures. We wouldn't expect to see a 1:1 relationship
between size and clustering. The age of the source may have allowed it
to expand despite a clustered environment, or conversely, perhaps we
see a young source which is expanding unconfined. Projection effects
may also hide the true size of the source.

\begin{figure}

	\plotfiddle{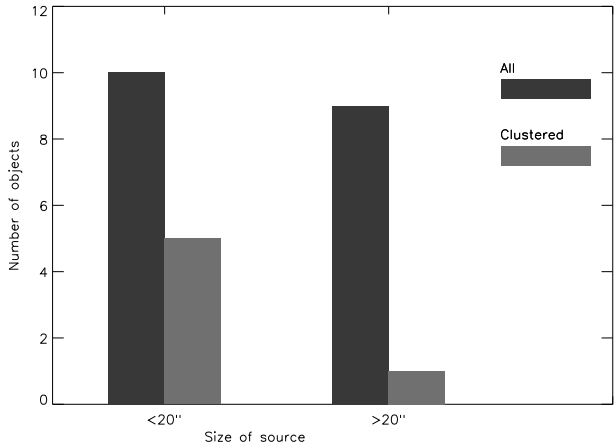}{4.5cm}{0}{50}{50}{-144}{-25}

	\vspace{0.5cm}

	\caption{ {\small Our sample divided into small and large
	steep spectrum sources. Clustered environments are more often
	associated with smaller radio structures.}}

\end{figure}

\section{Summary}

We have found that RLQs exist in a variety of environments from the
field through compact groups to rich clusters. In order to correctly
quantify the environments of these sources, multicolour information
must be used. There is evidence for a radio size - environment
relation with smaller sources located in richer systems. By selecting
small, powerful RLQs we can efficiently create a sample of distant
groups and clusters. We also have the ability to ``dial up'' clusters
at a specific redshift by looking at AGN.

\end{document}